# Snakes combine vertical and lateral bending to traverse uneven terrain

Qiyuan Fu[a], Henry C. Astley[b], Chen Li[a, *]

[a]Department of Mechanical Engineering, Johns Hopkins University, Baltimore, MD 21218, USA

[b]Department of Biology, University of Akron, Akron, OH 44325, USA

*Corresponding author. E-mail: [redacted]

**Keywords:** locomotion, limbless, complex terrain, terradynamics, contact

## Abstract

Terrestrial locomotion requires generating appropriate ground reaction forces which depend on substrate geometry and physical properties. The richness of positions and orientations of terrain features in the 3-D world gives limbless animals like snakes that can bend their body versatility to generate forces from different contact areas for propulsion. Despite many previous studies of how snakes use lateral body bending for propulsion on relatively flat surfaces with lateral contact points, little is known about whether and how much snakes use vertical body bending in combination with lateral bending in 3-D terrain. This lack had contributed to snake robots being inferior to animals in stability, efficiency, and versatility when traversing complex 3-D environments. Here, to begin to elucidate this, we studied how the generalist corn snake traversed an uneven arena of blocks of random height variation 5 times its body height. The animal traversed the uneven terrain with perfect stability by propagating 3-D bending down its body with little transverse motion (11° slip angle). Although the animal preferred moving through valleys with higher neighboring blocks, it did not prefer lateral bending. Among body-terrain contact regions that potentially provide propulsion, 52% were formed by vertical body bending and 48% by lateral bending. The combination of vertical and lateral bending may dramatically expand the sources of propulsive forces available to limbless locomotors by utilizing various asperities available in 3-D terrain. Direct measurements of contact forces are necessary to further understand how snakes coordinate 3-D bending along the entire body via sensory feedback to propel through 3-D terrain. These studies will open a path to new propulsive mechanisms for snake robots, potentially increasing the performance and versatility in 3-D terrain.

## INTRODUCTION

Unlike limbed animals, which typically generate support and propulsive forces at a few points in the environment with distinct anatomical structures (feet), elongate, limbless animals such as snakes can use their entire body to create a large number of contact points with the surrounding environment to move through (Gray and Lissmann, 1950). This enables the body of the snake to interact with the substrate at a wide range of local positions and orientations, and then modulate these interactions by altering force distribution among them. Lateral slithering motion has been the focus of much of the literature, focusing on either frictional interactions with smooth, rigid surfaces (Hirose, 1993; Hu et al., 2009), interactions with granular media (Schiebel et al., 2020a) and artificial turf (Gerald and Wass, 2019; Jayne and Bennett, 1989; Jayne and Bennett, 1990; Walton et al., 1990), or, most often, interactions with arrays of vertical structures replicating natural terrain objects such as plants and rocks on flat surfaces (Gray and Lissmann, 1950; Jayne, 1986; Jayne and Byrnes, 2015; Jayne et al., 2013; Kano et al., 2012; Schiebel et al., 2020b). These have inspired many snake robots to traverse similar environments using lateral bending (Gong et al., 2016; Hirose, 1993; Kano and Ishiguro, 2013; Sanfilippo et al., 2016; Wang et al., 2020). Lateral body bending in these scenarios relies on objects with special anisotropic properties or positions and orientations to press against. However, slithering is still commonly observed in a variety of 3-D terrains that lack such objects (Jayne and Byrnes, 2015; Jayne and Herrmann, 2011). This indicates that slithering snakes are able to generate propulsion by interacting with a wider range of terrain asperities using body deformation in all three dimensions.

Vertical bending during terrestrial snake locomotion is rarely studied. Previous work has focused on the use of vertical lifting to improve efficiency either by reducing frictional drag, such as in sidewinding (Marvi et al., 2014) and in sinus-lifting (Hirose, 1993), or by raising the body to reach higher surfaces (Gart et al., 2019). Recent studies have revealed that vertical body bending can be utilized by snakes to interact with terrains with significant height variations and generate propulsive forces to traverse them. For example, when the corn snake traverses a horizontal ladder lacking lateral contact points, it can generate substantial propulsive force and propulsive impulse by posteriorly propagating vertical waves with minimal lateral motion (Jurestovsky et al., 2021). The propulsive value of pure vertical bending was further confirmed by the success to traverse a similar terrain of a robophysical model replicating only the vertical bending (Jurestovsky et al., 2021). These recent observations in simplified environments with no lateral contact points suggested that vertical bending is promising for expanding the source of propulsion in natural 3-D environments by pressing against suitably oriented terrain asperities below the body, similar to lateral bending pushing against lateral contact points.

Understanding of these basic principles will have a major impact on snake robot locomotion in complex 3-D environments. Some previous snake robots traversed 3-D complex terrains using geometric gait designs that only apply to limited scenarios (Fu and Li, 2020; Jurestovsky et al., 2021; Lipkin et al., 2007; Takemori et al., 2018a; Tanaka and Tanaka, 2013). Some robots adapted simple cyclic gaits originally used on flat surfaces and passively conformed to vertical height variation of terrain by mechanical or control compliance (Takemori et al., 2018b; Travers et al., 2018; Wang et al., 2020). However, there is still a significant gap in snake robots' stability, efficiency, and versatility compared to animals in complex environments, in large part due to a lack of principled understanding of how to use vertical body bending to generate propulsion. In previous snake robots, vertical bending was used either to improve efficiency by reducing frictional drag (Marvi et al., 2014; Toyoshima and Matsuno, 2012) or to reach different terrain surfaces (Fu and Li, 2020; Lipkin et al., 2007; Takemori et al., 2018a; Takemori et al., 2018b; Tanaka and Tanaka, 2013; Wang et al., 2020). Only one snake robot used vertical bending to traverse a single cylindrical obstacle (Date and Takita, 2005), but that study assumed that there is no longitudinal friction, no lateral slipping, and no gravity. However, recent demonstration of a simple snake robot traversing a horizontal ladder in a similar fashion as snakes suggested that vertical bending can be used to generate propulsion over more complex 3-D terrain (Jurestovsky et al., 2021).

Inspired by these recent insights, here we take the next step in studying how snakes use 3-D body bending to move through the 3-D world. We hypothesize that, when both vertical and horizontal contact points are available, generalist snakes can use vertical body bending as frequently as lateral body bending to interact with and traverse 3-D terrain. As an initial step to test the hypothesis, we measured the kinematics of generalist corn snakes during their traversal of uneven terrain and analyzed their contact between their body and terrain surfaces. The uneven terrain allowed the animal to use both lateral and vertical bending for contact due to the variation of geometry in both vertical and horizontal directions. We evaluated performance by analyzing longitudinal vs. transverse motions and static stability. To estimate the relative contribution of lateral and vertical bending, we compared: (1) the number of body bends of each type contacting the terrain and (2) the number of horizontal and vertical bends which would potentially allow propulsive force generation.

## METHODS

### Animals

We used three captive-bred juvenile corn snakes [*Pantherophis guttatus* (Utiger et al., 2002)] purchased from an online vendor (Reptiles by Alex, Wichita, KS, USA). These species is a locomotor generalist commonly used in snake locomotion studies, and its natural habitats include substantial fine-scale terrain height variations such as forests and rocky hillsides (Burbrink, 2002; Conant and Collins, 1998). We housed snakes individually in 45.7 × 19.1 cm or 50.8 × 38.1 cm containers on a 12 h:12 h light:dark cycle at a temperature of 30 °C on the warm end and 25 °C on the cool end. Snakes were fed water and pinky mice. The snakes' full body length measured 82.2 ± 5.7 cm, and they weighed 165.0 ± 16.2 g. To measure length, we digitized dorsal view photos of each snake by tracing the body midline and scaling its length from pixels to centimeters (Astley et al., 2017). To quantify body tapering, we measured the cross-sectional height of the snakes at different locations by digitizing lateral view photos and interpolated values in between by fitting a quadratic polynomial to measured heights (Fig. S1A). All animal experiments were approved by and in compliance with The Johns Hopkins University Animal Care and Use Committee (protocol RE19A165).

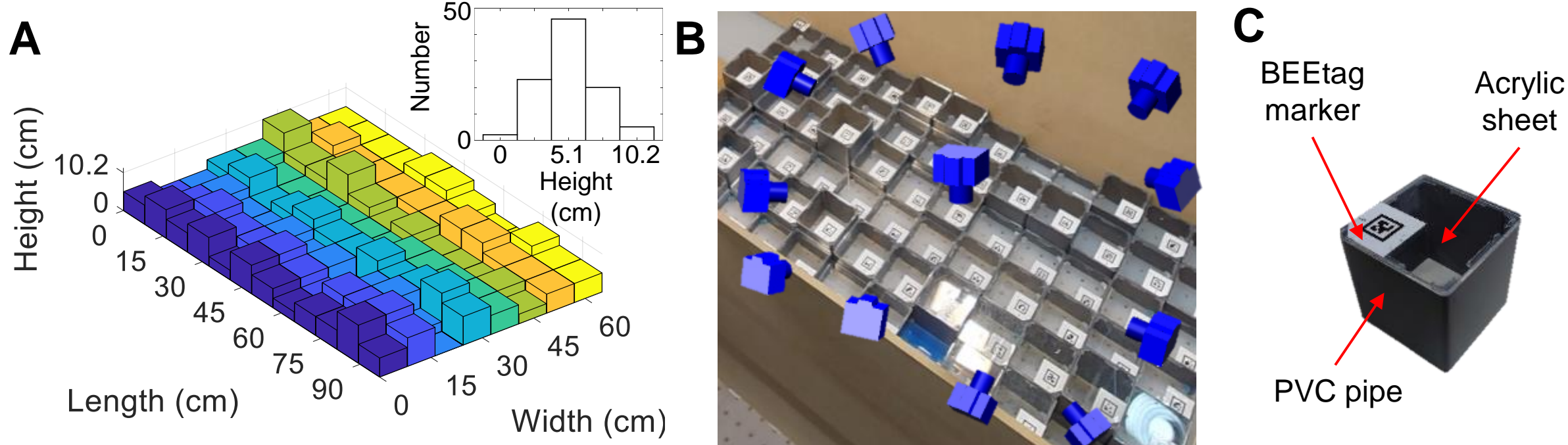

**Fig. 1. Experimental setup. (A)** Height distribution of terrain blocks. Inset shows histogram of block heights. **(B)** Photo of experimental setup. Blue shows twelve high-speed cameras used for 3-D reconstruction of markers. **(C)** Components of a terrain block.

**Uneven terrain arena**

We constructed a 65 cm wide, 97.5 cm long uneven terrain arena using 96 blocks in an 8 × 12 array (Fig. 1A-B). Each block has a horizontal footprint of 8.2 × 8.2 cm. Heights of all blocks follow a normal distribution (Fig. 1A) with a mean of 5.1 cm and a standard deviation of 1.7 cm. Positions and heights of all blocks were kept unchanged during all trials. Each block consisted of a rectangular PVC pipe (McMaster-Carr, Elmhurst, IL, USA) cut to desired height as side walls and a piece of acrylic sheet (McMaster-Carr, Elmhurst, IL, USA) laser cut to the same shape as the cross section of the PVC pipe as the top surface (Fig. 1C). The PVC pipe and the acrylic sheet were hot glued from the inside to keep the outer surface clean. Adjacent blocks were rigidly connected by 3-D printed clamps from bottom. To measure the 3-D positions of terrain blocks, we attached one 3.8 cm × 3.8 cm BEEtag marker (Crall et al., 2015) to each top surface of blocks and covered the marker with packaging tape (3M, Maplewood, MN, USA) to reduce friction. Three 61.5 cm tall wooden sheets were used as sidewalls to prevent the snakes from escaping and clamped together with the blocks using 3-D printed clamps from bottom.

We measured the kinetic friction coefficient between the snake body and the terrain blocks using a 3-axis force/torque sensor (Fig. S2A; ATI mini 40, Apex, NC, USA). The sensor measured normal force and friction while a snake was sliding against a plate rigidly connected to the sensor. We fit a line through origin to the normal force and friction collected during the middle 50% of time in each slide and calculated the friction coefficient from the slope. Each material that made up terrain blocks was used as the top surface of the plate for 5 measurements along each direction (forward, backward, left, and right) for each of the 3 animals. The kinetic friction coefficients between the snake body and acrylic, PVC, and packaging tape (covering the BEEtag markers), were $\mu = 0.32 \pm 0.05$, $0.28 \pm 0.07$, and $0.19 \pm 0.02$ (mean $\pm$ s.d.), respectively.

**Locomotion experiment protocol**

Snake locomotion was recorded using 12 high-speed cameras (Fig. 1B; Adimec, Eindhoven, The Netherlands) at 50 frames $s^{-1}$ with a resolution of $2592 \times 2048$ pixels. To illuminate the arena, two 500 W halogen lamps and two LED lamps were placed dorsally above the arena. The surface of the test area was heated to 32 °C during experiments. To calibrate the cameras for 3-D reconstruction, we made a $61 \times 66$ cm calibration grid out of DUPLO bricks (The Lego Group, Bilund, Denmark) and attached BEEtag markers (Crall et al., 2015) on it. We placed the calibration grid in the arena before experiments and recorded snapshots using the 12 cameras. To track the 3-D movements of the snake, we attached 10 to 12 lightweight (0.3 g) BEEtag markers along the dorsal side of the snake equally spaced ($\approx$ 6.6 cm) between neck and vent (Fig. S1B) using lightly adhesive tape ($0.4 \times 1.2$ cm).

The snake was kept in a hide near the test area at a temperature of 30 °C prior to experiments. We placed the snake on random locations inside the arena and encouraged it to traverse blocks by light tapping on the tail and a shaded shelter near the test area. A trial was ended if any part of the snake moved out of the test area or the snake stopped moving for more than 15 seconds. After each trial, the snake was removed from the test area, placed in the hide, and allowed to rest for 1 to 2 minutes.

After experiments, we tracked 2-D coordinates of the markers attached to the calibration grid and obtained intrinsic and extrinsic camera parameters using direct linear transformation (DLTdv5) (Hedrick, 2008). BEEtag markers attached to snakes and blocks were tracked in 2-D camera views and reconstructed for 3-D positions and orientations using custom MATLAB scripts (Crall et al., 2015; Hedrick, 2008). The geometry of terrain blocks was then reconstructed using measured dimensions and tracked positions and orientations.

**Continuous body 3-D kinematics interpolation**

To obtain continuous 3-D kinematics of snake body for contact analyses, we interpolated the midline of each section of body (both position and orientation) between adjacent markers by approximating it as an elastic rod subject to end constraints imposed by tracked markers (Fig. 2A, Fu et al., 2021). For interpolation, we used a rod with a constant radius throughout the body length for simplicity. Despite over-simplifying the biomechanics, the method has a higher interpolation accuracy (~50% less error) in both position and orientation than commonly used B-spline methods. The low position error (17% of body diameter on average (Fu et al., 2021)) of the interpolated midline enabled us to use it to reconstruct the surface of the snake body for contact analyses (see Contact analysis section below). When reconstructing the body surface (see below), we used measured body tapering instead of a constant radius for higher accuracy. Unless otherwise noted, we used the entire reconstructed midline in our analyses.

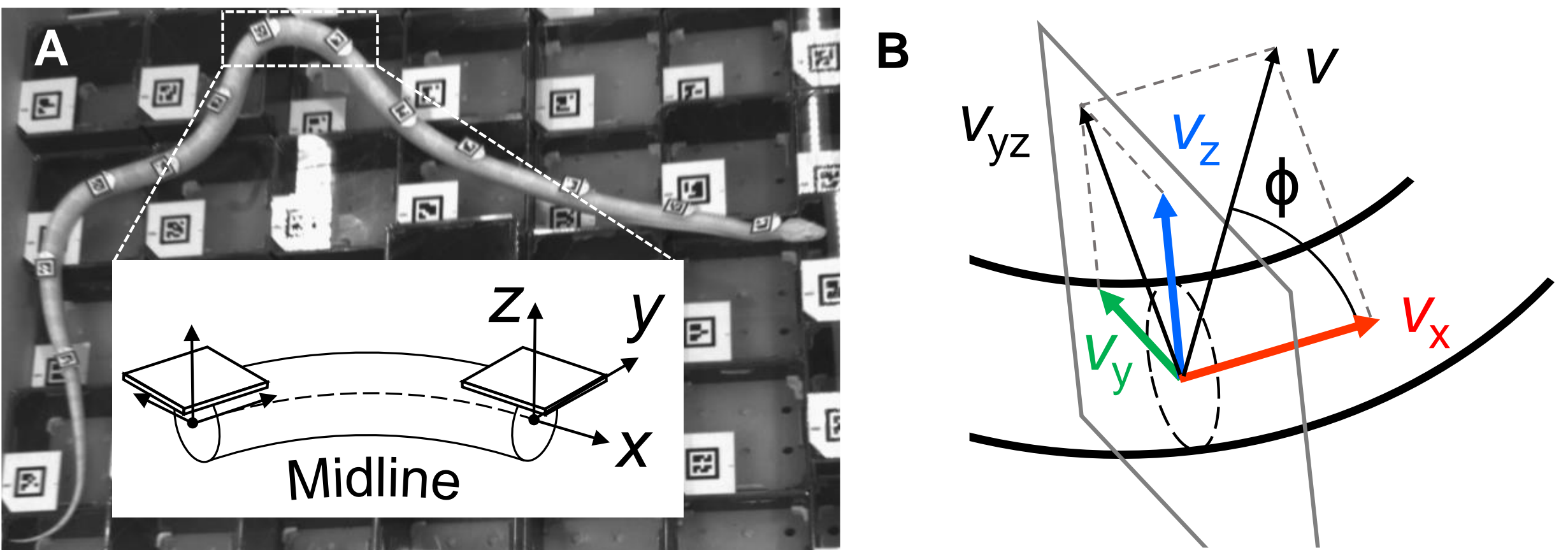

**Fig. 2. 3-D snake body reconstruction and definition of metrics to evaluate transverse motion. (A)** A reconstructed snake body segment between two adjacent markers and its midline (dashed). **(B)** Definition of local body velocity $v$, longitudinal velocity $v_x$, lateral velocity $v_y$, dorsoventral velocity $v_z$, transverse

velocity $v_{yz}$, and slip angle $\phi$. Dashed circle shows cross section and gray parallelogram shows local normal plane.

**Performance analysis**

To assess traversal performance, we calculated the accumulated distance traveled by the mid-body position (the mid point of the reconstructed midline) along its trajectory and the duration of travel using all video frames in each trial.

Occasionally, the snake stopped during traversal and most of the body remained static. In contrast, terrestrial slithering motion such as lateral undulation has forward longitudinal velocities throughout the body as the bending wave propagated down the body (Jayne, 2020). Thus, in all the following analyses, to avoid artificial bias from stopped body postures, we excluded video frames in which the average longitudinal velocity along the entire reconstructed body was small (< 0.125 cm $s^{-1}$) (Fig. S3A; 6.4% of all the video frames).

To assess how effectively the corn snake used substrate irregularities in the uneven terrain arena to generate propulsion, we calculated longitudinal and transverse velocities and slip angle. During slithering locomotion, posteriorly propagating bends of the body generate reaction forces against the substrate to propel the snake forwards (Gray and Lissmann, 1950; Mosauer, 1932). On flat, rigid planes (rarely found in nature), frictional anisotropy of the scales allows generation of forward forces, but with high slipping and low speeds (Alben, 2013; Hu et al., 2009; Schiebel et al., 2020a). However, with the presence of geometric asperities against which body bends can push without constant yielding (e.g., rocks, plants, sufficiently large piles of sand), snakes will show minimal slip as if moving in a virtual tube with greatly increased speed (Gray, 1946; Gray and Lissmann, 1950; Jayne, 1986; Kelley et al., 1997; Mosauer, 1932; Schiebel et al., 2020a; Schiebel et al., 2020b).

For each infinitesimal body segment of the reconstructed midline, we calculated longitudinal velocity as the velocity component parallel to the local body segment, $v_x = \vec{v} \bullet \vec{T}$, and transverse velocity as the velocity component perpendicular to it, $v_{yz} = |\vec{v} - \vec{v} \bullet \vec{T}|$ (Fig. 2B). We calculated slip angle as the angle between local body velocity and local body tangent in 3-D ($\phi = \cos^{-1}(\vec{v} \cdot \vec{T}/|\vec{v}|)$, which ranges from 0º to 180º, Fig. 2B) (Sharpe et al., 2015), which measures how well the body stays within a virtual tube as it progresses. Perfectly progressing forward in a virtual tube results in a slip angle of 0º, whereas no progress or backward progress in it results in a slip angle of 90º or 90-180º, respectively.

We calculated longitudinal and transverse velocities and slip angles for all body segments regardless of contact conditions (next section) to quantify how well the overall kinematics matches ideal slithering motion within a virtual tube. We also calculated these three metrics only for the body segments in contact with the terrain, which can alter contact forces and affect propulsion performance more than that of body segments suspended in the air.

To test whether the anterior end of the snake moved transversely more than the other part of the body, we divided the snake body into two parts: the 10% of body segments closest to the nose and the other 90%, defined as the anterior region and the main body region, respectively (Fig. S1A).

To assess whether the snake tended to move on lower blocks (as if moving in a valley), we compared average height of blocks directly below the snake body ($h_{\text{below}}$) and that of neighboring blocks ($h_{\text{neighbor}}$) (see Data averaging below). Neighboring blocks were defined as blocks that were adjacent to blocks directly below the snake body. A neighboring block may contact the lateral sides of the snake body under this definition. We also calculated the percentage of video frames in each trial when the average height of all blocks directly below the body was smaller than that of neighboring blocks.

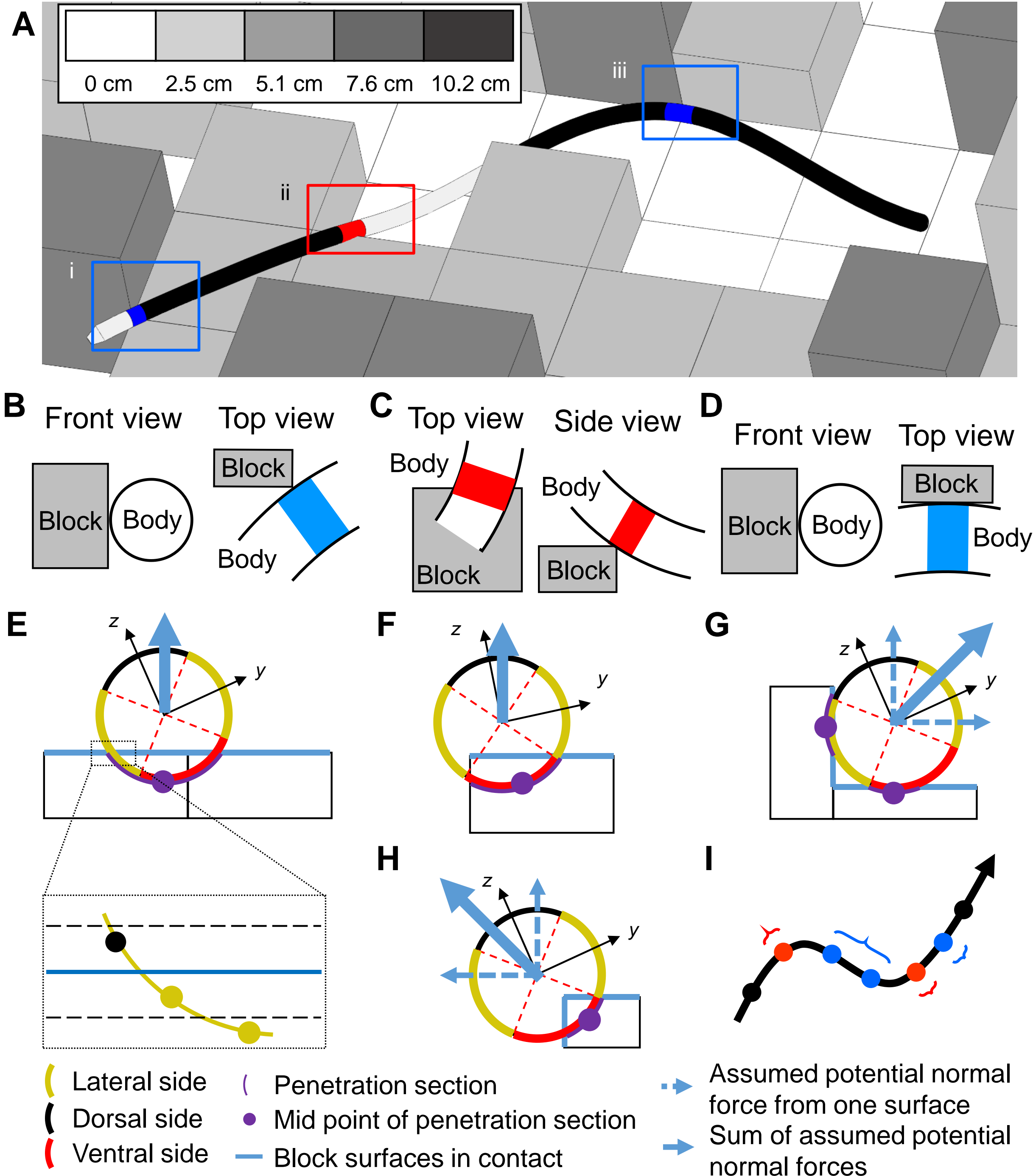

**Fig. 3. Examples to determine contact types and contact regions. (A)** An example of contact types of different body regions. Blue (lateral contacting), red (vertical contacting), black (supported), and white (suspended) indicate different body-terrain contact types. Three insets show representative cases of body-

terrain contact: (i) Body is laterally contacting a vertical edge. (ii) Body is vertically contacting a horizontal edge. (iii) Body is laterally contacting a vertical wall. **(B-D)** Different views of case (i-iii) in (A). **(E-H)** Examples to show how contact types are determined by where cross-sectional outline (circle) contacts terrain blocks (boxes). (E) A supported body segment that only contacts horizontal surfaces. Inset shows identification of surfaces a body segment is contacting. Black dashed lines show range in which a sample point is considered in contact with terrain surface. Yellow and black points show sample points penetrating and outside the terrain block. (F) A special case of supported body segment that is considered to only contact a horizontal surface and vertical surface of the block below the segment is ignored because otherwise body segments sitting on a horizontal edge are falsely classified as vertical contact segments. (G) A lateral contact body segment that contacts one vertical surface on its lateral side. (H) A vertical contact body segment that contacts one vertical surface only with its ventral side. In (E-H), assumed kinetic friction force is perpendicular to sum of assumed potential normal forces and opposite to local body velocity. **(I)** An example to determine contact regions. Black line shows body midline. Points show body segments and brackets show contact regions.

**Contact analysis**

To classify contact types of different body parts (Fig. 3A), we first determined contact between the snake body and the terrain surfaces. We sampled 200 locations evenly along the reconstructed midline and 24 points on the circumference of the cross-sectional outline (assumed to be circular) of each sampled body segment, resulting in a total of 4800 points on the reconstructed snake body surface. Each cross-sectional outline was radially expanded outward from the reconstructed midline by the fitted local body radius to account for effects of body tapering. Collision detection between these sample points and each reconstructed terrain block was performed to locate contact points using the GJK algorithm (Sheen, 2021), a common algorithm to determine collision between convex objects. Only blocks directly below and neighboring blocks were included for collision detection to save computation time.

To identify terrain surfaces that a body segment was contacting, we checked the distances between each point on the sampled outline where the outline started to penetrate blocks (Fig. 3E inset, top yellow point) and each face of the block that this point was penetrating (Fig. 3E, blue solid line). Faces obstructed by other blocks were not considered (Fig. 3E). Vertical surfaces of the block directly below the body segment were not considered because otherwise the body segment sitting on a horizontal surface along an edge was falsely classified as contacting the vertical surface directly below it (Fig. 3F).

To check whether a body segment contacted terrain on the lateral sides or on the ventral side, we divided the outline of each body segment into four sections of equal length (one ventral, one dorsal, and two laterals; Fig. 3E-H, red, black, and yellow arcs, respectively) and checked into which section the midpoint (Fig. 3E-H, purple point) of each penetration section (Fig. 3E-H, purple arc) fell.

By checking which terrain surfaces the body segment contacted and on which sides of the body segment the contact happened, we classified sampled infinitesimal body segments into 4 types: (1) Suspended (Fig. 3A, white): the segment was not contacting the terrain. (2) Supported (Fig. 3A, black; Fig. 3E): the segment was contacting horizontal terrain surfaces only. However, if the segment was also contacting vertical surfaces, it was classified as either (3) or (4) below. (3) Lateral contact (Fig. 3A, blue; Fig. 3B, D, G): the segment was contacting vertical walls with a lateral side. This includes laterally contacting a vertical wall (Fig. 3A, right blue) or a vertical edge connecting two vertical walls (Fig. 3A, left blue). (4) Vertical contact (Fig. 3A, red; Fig. 3C, H): the segment was contacting vertical walls only with its ventral side. This includes contacting a horizontal edge (Fig. 3A, red) or vertices connecting multiple horizontal edges with its ventral side. Contact types of the infinitesimal body segments not sampled for collision detection were interpolated using the values of the nearest sampled body segment. Results of this classification were visually examined by color-coding the reconstructed midline accordingly in camera videos (Fig. 4A) and flattened sagittal views (Fig. 4B).

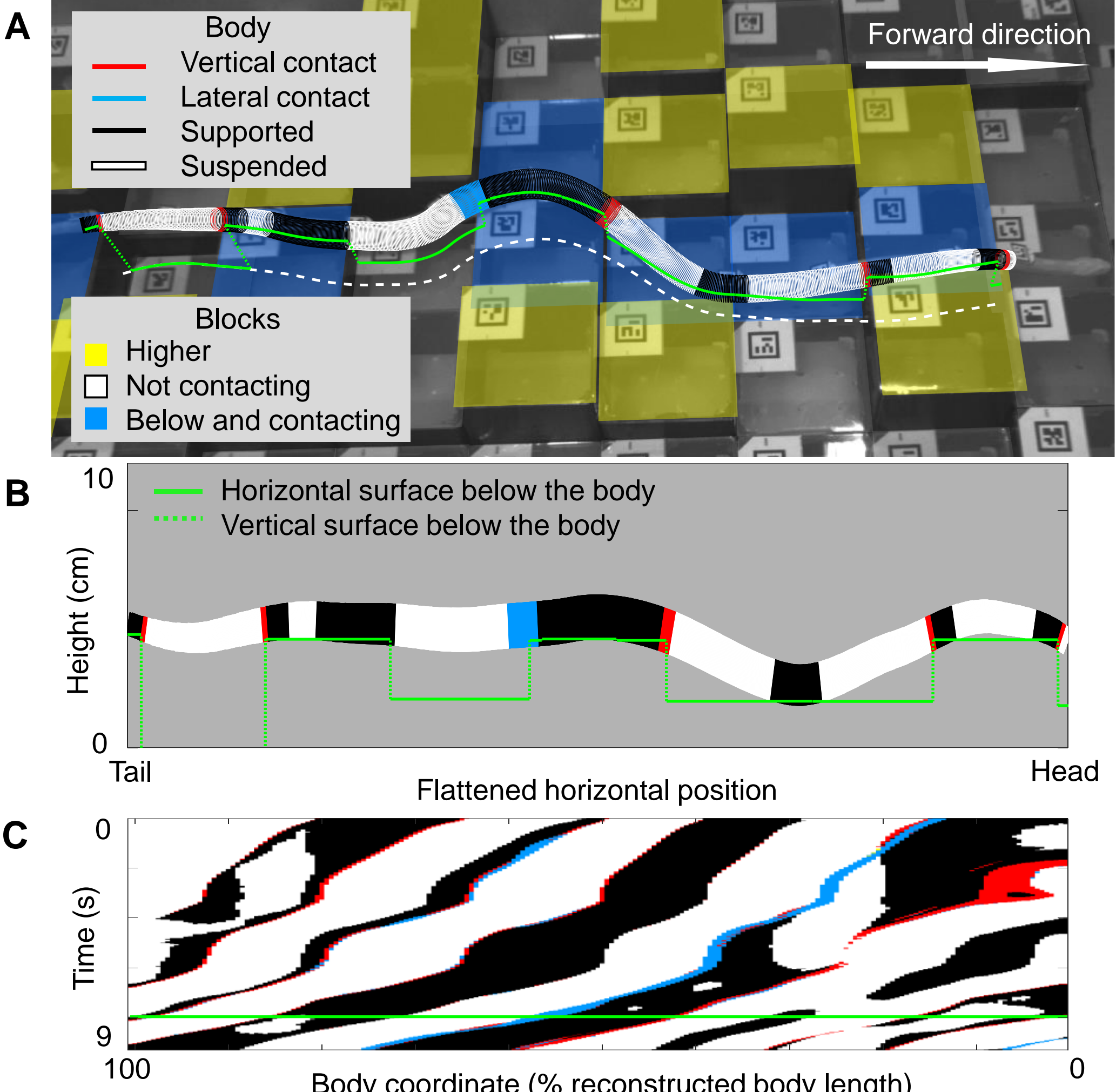

**Fig. 4. Contact types of snake body at a representative moment. (A)** A representative snapshot of snake traversing uneven terrain in which the snake is moving from left to right. Colors of reconstructed snake body surface show body-terrain contact types. Colors of reconstructed top surfaces of terrain blocks show block heights with respect to nearby snake body. Green solid, green dotted, and white dashed curves indicate intersections between the curved body sagittal plane and terrain top surfaces, terrain side surfaces, and ground horizontal plane, respectively. Thus, the white dashed line is a top-down view, albeit viewed from an oblique perspective. **(B)** Flattened sagittal view of reconstructed snake body and its intersections with the different surfaces in (A). Flattened horizontal position is the accumulated horizontal distance to the head in the curved body sagittal plane. **(C)** A representative spatiotemporal profile of contact types as a function of body coordinate and time. Body coordinate is defined as ratio of cumulative length along the body from head to reconstructed body length. Green line corresponds to snapshot in (A). This representative trial does not contain phases in which the snake stops (i.e., video frames in which the average longitudinal velocity along the entire reconstructed body is small (< 0.125 cm $s^{-1}$)). Other trials may contain such phases but these video frames are excluded from most analyses (see Performance analysis). In (B, C), body coordinate starts from 0% at the most anterior marker and ends at 100% at the most posterior marker. See Movie 1 for a representative video.

To quantify vertical and lateral bending used by animals for contact, we counted the number of body sections in contact with the terrain (referred to as contact regions hereafter) formed by continuous body segments with lateral or vertical contact. A contact region was counted for each continuous section of the reconstructed midline that was made up only by body segments of one contact type (Fig. 3I). Each contact region may have variable length and shape. We counted the number of contact regions instead of the number of body segments because a vertical contact region contacting an edge (Fig. 4A, red) often appeared with fewer body segments than a lateral contact region contacting a surface (Fig. 4A, blue).

Because we could not measure forces directly, we used simple assumptions to infer the likely terrain reaction force directions considering the rectangular geometry of the arena. Regardless of force magnitude, the reaction force against any surface must be the sum of the outward force normal to the surface and the frictional force along the surface opposite to local body velocity, which is proportional to normal force by the coefficient of friction. We only considered kinetic friction because of a lack of force measurements required to determine the direction of static friction. Thus, we did not consider friction from body segments that had a small total velocity (< 0.3 cm $s^{-1}$) (Fig. S3C; 3.3% of all the body segments in al the video frames pooled from all trials of all animals). If a body segment was contacting multiple surfaces (e.g., two vertical surfaces that met at a vertical edge), we assumed that the potential normal force generated from it (Fig. 3G, H, blue solid arrows) was along the sum of the normal vectors of all these surfaces (Fig. 3G, H, blue dashed arrows). We assumed a body segment to be generating propulsion if the sum of potential normal force and potential kinetic friction force generated from it has a positive projection in the forward motion of overall body movement (i.e., if the total potential force direction and the instantaneous center of mass velocity formed an angle smaller than 90°) in the top view.

We considered a contact region to be potentially propulsive if any of the body segment within it was assessed to be propulsive, assuming that the animal can redistribute force to use any potentially propulsive body segment within a contact region. Given the lack of direct force measurement, we could not assume that contact surface area is positively correlated with the force exerted by each body region, as small contact points may generate high forces and large ones may generate lower forces. This has never been assessed in limbless locomotion. Instead, we assumed that each contact region had the same potential to propel the body and counted the number of potentially propulsive contact regions. We did not consider video frames (Fig. S3B; 7% of all frames) with slow instantaneous center of mass velocities (< 0.3 cm $s^{-1}$) when evaluating likely propulsive body segments to mitigate tracking noise.

To test whether the snake was stable, we first estimated center of mass position by averaging positions of all interpolated body segments (96% of full body volume as estimated from tapering data) weighted by cross-sectional area in each video frame. Then, we checked whether center of mass projection onto the horizontal plane (Fig. S4, red circle) fell into the support polygon (Fig. S4, purple polygon), a 2-D convex hull enclosing the projection of body segments in contact with horizontal surfaces (including their edges) of the terrain (Gart et al., 2019). We calculated static stability performance for each trial by dividing the number of video frames in which the snake was statically stable with the total number of video frames.

**Sample size**

We performed experiments using three snakes ($N = 3$) with 18 trials for each animal. After rejecting trials with large reconstruction errors because of loss of tracking of occluded markers for a long time, 8, 9, and 13 trials remained for the three individuals, respectively, resulting in a total of $n = 30$ accepted trials. Video frames in which part of the body was not reconstructed because markers were occluded by blocks (0.08% of all video frames) were excluded from statistical tests.

**Data averaging**

*1. Metrics directly calculated for each body segment or each terrain block*

For each trial, we first averaged these metrics spatially in each video frame and then averaged them temporally across all video frames (excluding frames in which the snake stopped). These metrics include: (1) average longitudinal and transverse velocities and slip angle of the entire reconstructed body; (2) average longitudinal and transverse velocities and slip angle of the body sections in contact with the terrain; (3) slip angle of the anterior region; (4) slip angle of the main body region.

To obtain height difference between the blocks directly below snake body and the neighboring blocks, we first calculated the average height of all blocks below the body and that of all neighboring blocks, because the number of these two types of blocks were often unequal. We then calculated their difference in each video frame and averaged them temporally across all video frames (excluding frames in which the snake stopped).

*2. Metrics directly calculated for each video frame*

For each trial, we averaged these metrics temporally across all video frames. These measurements include: (1) accumulated distance traveled by the mid-body position along its trajectory; (2) duration of travel; (3) number of vertical and lateral contact regions; (4) number of potentially propulsive vertical and lateral contact regions. Video frames in which the snake stopped (< 0.125 cm $s^{-1}$) were excluded for (3) and (4).

*3. Metrics directly calculated for each trial*

No averaging was necessary for these metrics. These metrics include: (1) static stability performance; (2) percentage of video frames in which the average height of the blocks under the animal was smaller than that of the neighboring blocks; (3) percentage of video frames in which all the potentially propulsive regions were purely vertical contact region or purely lateral contact regions.

By performing a random-effect ANOVA with individual as a random factor, we verified that the intra-subject variance was larger than the between-subject variance (Leger and Didrichsons, 1994). Thus, we pooled all trials of all individuals. Finally, for metrics in categories 1 and 2, we calculated means and standard deviations of all trials. For metrics in category 3, we calculated median and interquartile range of all trials or plotted histogram of all trials in the case where the data had a non-normal distribution.

**Statistical tests**

To test whether two paired measurements within a trial differed consistently, we performed paired *t*-tests pooling all trials from all animals. These paired measurements include heights of blocks directly below the body versus neighboring blocks, transverse versus longitudinal velocity, slip angle of the anterior part versus the rest of the body, the number of lateral versus vertical contact regions, and the number of potentially propulsive lateral versus vertical contact regions.

To test whether kinetic friction coefficient between the snake body and a surface material differed along different directions, for each individual and each surface material, we performed an ANOVA followed by a Tukey's honestly significant difference (HSD) test with sliding direction as an independent variable and kinetic friction coefficient as a dependent variable.

All the statistical tests followed (McDonald, 2014) and were performed using JMP Pro 15 (SAS Institute, Cary, NC, USA).

## RESULTS

**Traversal behavior**

The animal traversed the uneven terrain by propagating 3-D bending down the body with little transverse motion out of the virtual tube (Fig. 5A, Movie 2), similar to prior studies of snakes moving in heterogeneous terrain such as artificial turf and surfaces with arrays of vertical structures (Jayne, 1986; Kano et al., 2012; Schiebel et al., 2020b). However, the snake often did not form clear periodic wave forms during traversal. The animal's mid-body position (midway from the head and from the tail) traveled along its trajectory an accumulated distance of $52.8 \pm 23.2$ cm ($0.64 \pm 0.28$ body length) within $15.5 \pm 6.4$ s. For the entire reconstructed body, the average longitudinal velocity was $3.9 \pm 1.4$ cm $s^{-1}$ ($4.8 \pm 1.7$% body length $s^{-1}$) and the highest average longitudinal velocity in all video frames from all trials was 18.0 cm $s^{-1}$ (22% body length $s^{-1}$). The highest velocity occurred after the snake was tapped lightly on its tail. Slip angle of the entire reconstructed body (which is in 3-D) was small (12° ± 3°), around 41% of that when corn snakes move on a rigid, smooth substrate (28° (Schiebel et al., 2020b). Transverse velocity was only 16% of longitudinal velocity (0.6 vs. 3.9 cm $s^{-1}$; Fig. 5B, C; $t(29) = 15.52$, $P < 0.0001$, paired *t*-test). For the body sections in contact with the terrain, slip angle was 10° ± 3°, and transverse velocity was only 14% of longitudinal velocity (0.5 vs. 3.8 cm $s^{-1}$; Fig. 5C; $t(29) = 15.23$, $P < 0.0001$, paired *t*-test).

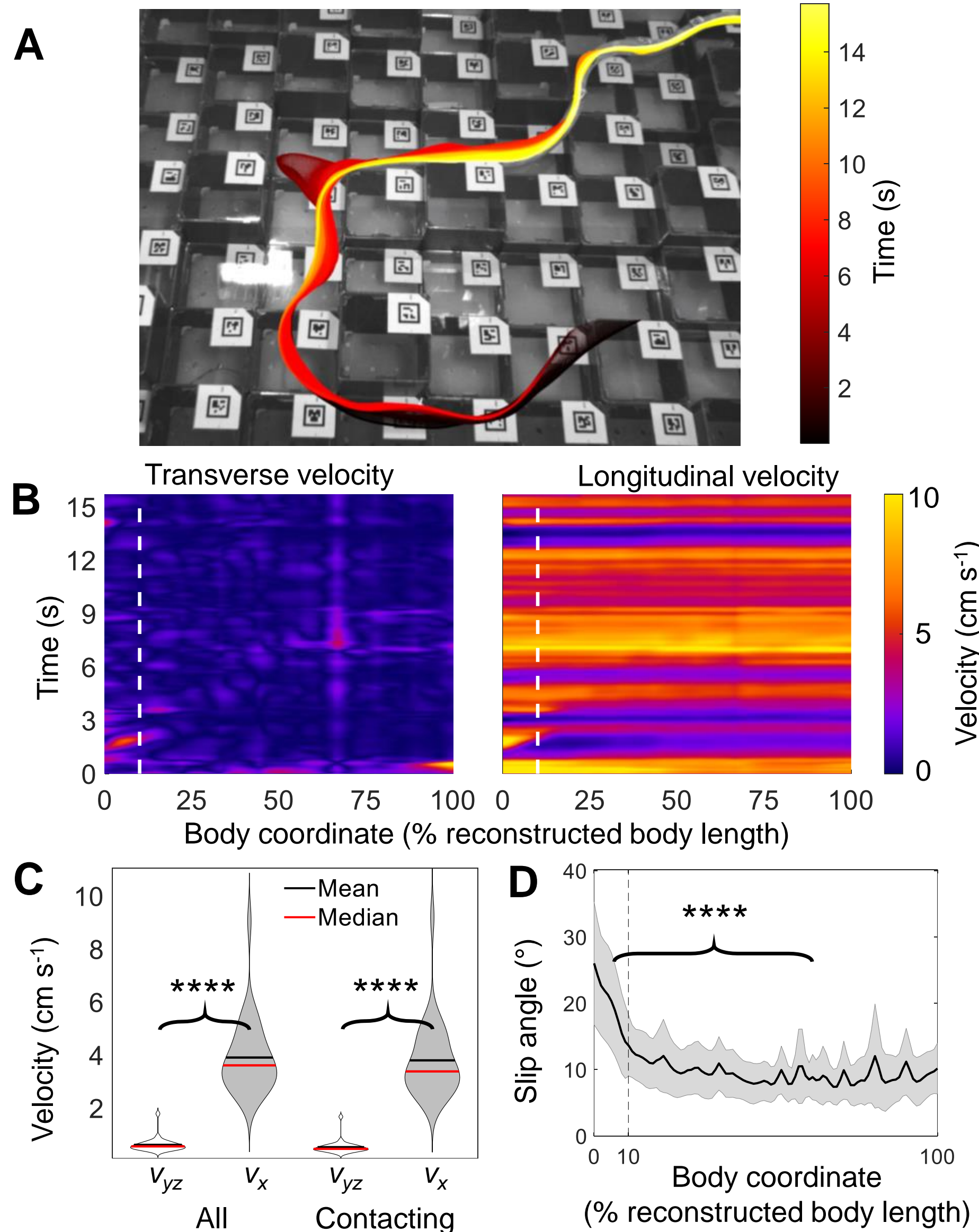

**Fig. 5. Representative snapshot showing little transverse motion. (A)** Representative snapshot of snake with reconstructed midline overlaid at different time instances during traversal of uneven terrain. Midline

color changes from black to light yellow with elapse of time from start to end of traversal. See Movie 2 for a representative video. **(B)** Spatiotemporal profiles of transverse velocity (left) and longitudinal velocity (right) as functions of body coordinate and time in the same trial in (A). Vertical white dashed lines indicate division between the anterior region and main body region. This representative trial does not contain phases in which the snake stops (i.e., video frames in which the average longitudinal velocity along the entire reconstructed body is small ($< 0.125$ cm $s^{-1}$)). Other trials may contain such phases but these video frames are excluded from most analyses (see Performance analysis in METHODS). Body coordinate starts from 0% at the most anterior marker and ends at 100% at the most posterior marker. **(C)** Transverse ($v_{yz}$, white) and longitudinal ($v_x$, gray) velocity of all the reconstructed body sections (left) and of the body sections in contact with the terrain (right). Data is shown using violin plots. Black and red lines show mean and median. Local width of graph is proportional to smoothed probability density of data along the y-axis (Hoffmann, 2021). **(D)** Slip angle along the body. Black curve and shaded area show mean ± s.d. Gray dashed lines indicate division between the anterior region and the main body region. Brackets and asterisks represent a significant difference (****$P < 0.0001$, paired $t$-test). $N = 3$ individuals, $n = 30$ trials.

The anterior region moved out of the virtual tube more than the main body region (Fig. 5D; $t(29) = 9.35$, $P < 0.0001$, paired $t$-test), with a twice larger slip angle (anterior: 19.70° ± 6.4° vs. main: 9.4° ± 2.3°, respectively). Video observation indicated that this may result from the exploration behavior of the head which occurred in all trials (see Movie 1 for an example). The anterior region frequently moved laterally or dorsoventrally as if exploring and selecting a path, while the main body region mostly followed the path of the anterior points.

**Body-terrain contact**

The animal tended to move through lower "valleys" surrounded by higher neighboring blocks. Average height of blocks directly below snake body in all trials was 1.2 cm (147 % body height) smaller than that of neighboring blocks on average (Fig. 6A; $t(29) = 6.66$, $P < 0.0001$, paired $t$-test). The average height of the blocks under the animal was smaller than that of the neighboring blocks in 100% (median) of the video frames across all trials of all individuals (Fig. S5A).

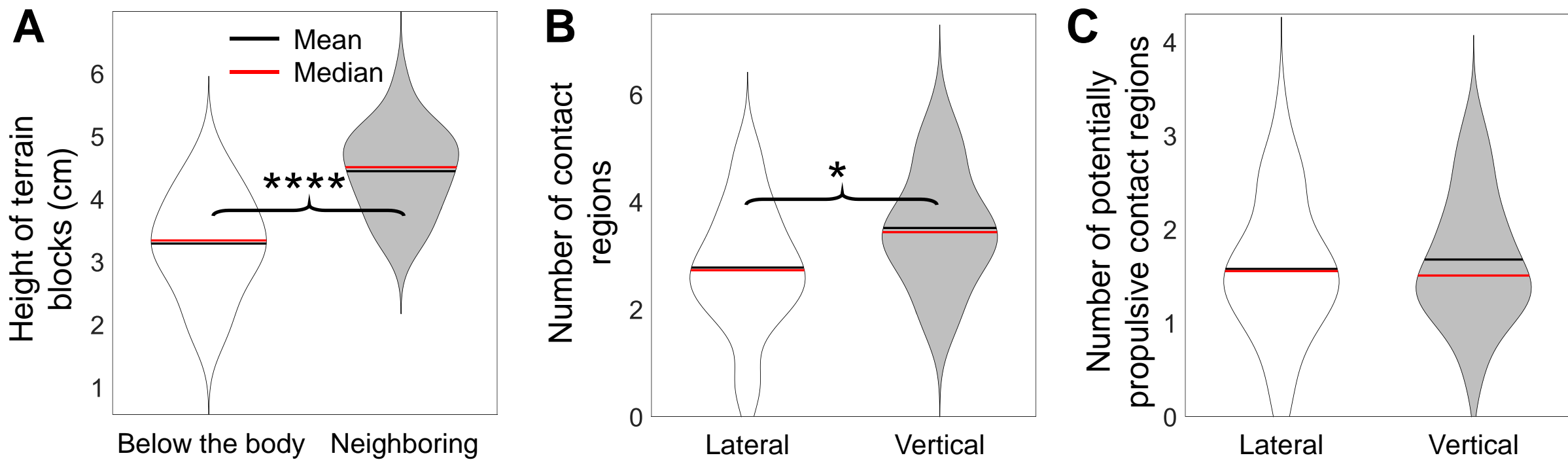

**Fig. 6. Quantification of movement in a valley and comparison of contact types. (A)** Comparison of height of terrain blocks directly below snake body and height of neighboring blocks. **(B)** Comparison of number of lateral and vertical contact regions. **(C)** Comparison of number of lateral and vertical contact regions that are likely propulsive. Data are shown using violin plots. Black and red lines show mean and median, respectively. Local width of graph is proportional to the probability density of data along the y-axis (Hoffmann, 2021). Brackets and asterisks represent a significant difference (****$P < 0.0001$, *$P < 0.05$, paired *t*-test). $N = 3$ individuals, $n = 30$ trials.

Despite this tendency, lateral contact with higher blocks was not utilized by the animal more frequently than vertical contact during traversal (Fig. 6B). The number of vertical contact regions in all trials was statistically larger than the number of lateral contact regions (3.5 vs. 2.8; Fig. 6B; $t(29) = 2.66$, $P < 0.05$, paired *t*-test). In 3 out of all 30 trials, the number of vertical contact regions along the body during traversal was more than 4 times that of lateral contact regions (see Movie 1 for an example). However, the maximum ratio of the number of lateral contact regions with respect to that of vertical contact regions in all trials was only 2.2.

The animal used lateral and vertical bending similarly often to form potentially propulsive contact regions, with $1.6 \pm 0.7$ potentially propulsive lateral and $1.7 \pm 0.7$ vertical contact regions across all trials, respectively (Fig. 6C; $t(29) = 0.52$, $P = 0.61$, paired *t*-test). In 8% (median; with interquartile range of 1% to 23%) or 7% (median; with interquartile range of 1% to 20%) of video frames in each trial, all the potentially propulsive regions were purely vertical contact region or purely lateral contact regions, respectively.

The animal traversed the uneven terrain with perfect stability, with the center of mass falling within the support polygon formed by body segments in contact with horizontal surfaces all the time (Fig. S5B; median across all trials from all individuals: 100%). Video observation showed that the few video frames estimated to be unstable resulted from the underestimation of support polygon. This is because we could not interpolate the body shape beyond the most anterior and the most posterior markers.

## DISCUSSION

### Contribution and implications

Our observations and quantification of types of body-terrain contact supported the hypothesis that vertical bending is used by generalist snakes to push terrain as frequently as lateral body bending during traversal of uneven terrain. We observed that supported contact regions sometimes ($0.9 \pm 0.8$ regions across all trials) experienced kinetic friction pointing toward the center of mass velocity. These supported regions occurred when the snake was bending in a U shape in the horizontal plane. When the center of mass velocity pointed towards the head, the rear part of the U-shaped body had local velocities that were opposite to center of mass velocity and generated such friction. However, it remains to be studied whether the supported regions were actively controlled to generate propulsion using static or kinetic friction.

The combination of lateral and vertical bending in 3-D may drastically expand the range of natural surfaces available for propulsion generation in all but the smoothest environments (Gart et al., 2019; Jurestovsky et al., 2021), by allowing each part of the entire body to adaptively push against its nearby terrain surfaces. This expanded range would allow snakes to better maintain propulsive forces to overcome frictional resistance continuously. This is important because, unlike legged locomotion which is affected significantly by inertial forces (except for tiny animals like ants (Clifton et al., 2020; Hooper, 2012) and mites (Weihmann et al., 2015)) and allows momentary loss of propulsive forces during continuous movement, terrestrial limbless slithering is mostly dominated by frictional forces and stops immediately after losing propulsion (Chong et al., 2021; Hu et al., 2009). The expanded range may also give snakes more redundancy to adjust distribution of contact forces to improve propulsion, stability, maneuverability, and efficiency, contributing to snake's locomotor versatility.

One potential advantage of vertical bending over lateral bending in providing propulsion is that obtaining vertical contact points is relatively easier in certain environments that have a small density of asperities large enough for lateral contact but substantial height variation over the entire body length, such as when snakes move over horizontal branches (Jurestovsky et al., 2021), travel down large boulders or move inside vertically bent tunnels. In such environments, the slender, elongated body has a high probability to ventrally contact terrain structures with height differences that are available for propulsion using vertical bending. Gravity pulls part of the body down to overcome frictional resistance, and continuous bending propagation allows such a process to continue at posterior body sections as long as height differences exist. The lower the belly friction is, the smaller slope angle is needed for the gravity to overcome frictional resistance, and thus a greater fraction of environmental surfaces can be utilized by using vertical bending to generate propulsion. However, to contact large asperities using lateral bending, it may need to reach laterally for a long distance before contacting such structures. Another potential advantage of vertical bending for propulsion is that force components along undesired directions can be easier to balance for stability by gravitational force. By contrast, lateral bending to contact vertical structures may be difficult to perform continuously without large yawing or lateral slipping unless there is a sufficient density of suitable asperities on both sides of the body (Gans, 1962).

Our results showed that snakes can seamlessly combine vertical and lateral body bending to generate propulsion in a three-dimensional complex environment. Combined with a recent study showing that snakes can generate propulsive force from vertical bending (Jurestovsky et al., 2021) much like lateral bending (Gray and Lissmann, 1950), this suggests that lateral undulation (Jayne, 2020) and vertical undulation are merely special cases induced by vertically and laterally homogenous environments, respectively, of an inherently three-dimensional behavior. This raises the question of whether such slithering using 3-D body bending propagation should be classified as a general mode of limbless locomotion (Jayne, 2020).

**Limitation and future work**

Our study is only an initial step towards understanding how snakes should combine vertical and lateral body bending to push against and move through the 3-D world. To further confirm our hypothesis, we must further measure 3-D contact forces between the body and terrain. This is challenging because high-fidelity commercial 3-D force sensors are expensive (Han et al., 2021; Jurestovsky et al., 2021) whereas low cost, customizable force sensors are typically 2-D and have low fidelity (Fu and Li, 2021; Kalantari et al., 2012; Liljebäck et al., 2012; Shimojo et al., 2007; Sundaram et al., 2019). We are developing a proof-of-concept custom 3-D force sensor achieving high fidelity with a relatively low cost. We still need to create a complex 3-D terrain platform with these force sensors embedded and controlled by data acquisition systems to ensure a sufficient sampling frequency.

To further understand how animals control 3-D body bending, measurements of muscle activity and neural signals are needed to answer the following questions: Does the animal also actively control scale (Marvi and Hu, 2012) and skin (Newman and Jayne, 2018) movement during this process? Is propulsion generation using vertical body bending actively controlled throughout the body by the propagation of epaxial muscular activation (Jayne, 1988; Moon and Gans, 1998), or can snakes use gravity (Jorgensen and Jayne, 2017) to facilitate this process? Although the 3-D shape did not change much as it was passed down the body, the animal often did not form clear alternating left and right bends (a hallmark of lateral undulation (Jayne, 2020)) or alternating up and down bends. Does bending in 3-D require using muscles differently from that during lateral undulation in terrestrial (Jayne, 1988; Moon and Gans, 1998) and arboreal (Astley and Jayne, 2009) environments?

In addition, future studies should test how generalist snakes modify their 3-D body bending to adapt to terrain properties change. Does their preference of using lateral or vertical bending depend on their habitat terrain properties, such as the spatial density of lateral and vertical push points available (Jayne and Herrmann, 2011; Schiebel et al., 2020b; Sponberg and Full, 2008) or friction (Alben, 2013; Gray, 1946; Marvi and Hu, 2012; Zhang et al., 2021)?

More broadly, it remains to be discovered how generally the combination of lateral and vertical bending is utilized by other snake species and other limbless clades in various terrains for propulsion. Aside from the locomotor generalist corn snakes studied here, other snakes including boas, pythons, sunbeam snakes, and many other colubrids have been observed to use similar movements (H. C. Astley, personal observation). Other limbless clades such as worms (Dorgan, 2015; Kwon et al., 2013) and fish (Ekeberg et al., 1995; Gidmark et al., 2011; Tatom-Naecker and Westneat, 2018) can also bend the body in three dimensions, but previous studies had focused on homogeneous environments like agar, gelatin, or sand until very recently (Pierce et al., 2021). Future studies will test this and reveal how the effectiveness of this strategy depends on the specie's specific neuromechanics, such as body bending capacity (Jurestovsky et al., 2020; Kelley et al., 1997), mechanical (Donatelli et al., 2017) and controlled (Marvi and Hu, 2012; Newman and Jayne, 2018) local compliance, muscular torque capability in each direction (Astley, 2020; Long Jr, 1998), and sensing and neural control capacity (Sulston et al., 1983). This strategy's effectiveness is also likely affected by habitat terrain properties, such as push point density (Majmudar et al., 2012), friction (Dorgan et al., 2013), deformability (Gu et al., 2017), and heterogeneity (Mitchell and Soga, 2005).

These comparative studies will provide insight into their habitat use and the links between habitat, morphologies, biomechanics, and performance within and between species. For instance, unlike limbed animals that generate propulsion by stepping on surfaces with slope grades shallower than the coefficient of friction (i.e., operate within the friction cone (Klein and Kittivatcharapong, 1990)) to avoid slipping, limbless animals may prefer utilizing surfaces with slope grades steeper than the coefficient of friction (i.e., operate outside the friction cone) in order to slither through. This would allow limbless animals to shelter in complex, confined environments cluttered with heterogeneous structures that are challenging for limbed animals, which may explain dozens of independent evolutionary convergences of limbless species (Gans, 1986).

The combination of lateral and vertical bending in 3-D should also be used by snake robots to fully exploit environmental surfaces with various positions and orientations for propulsion and stability. The wide range of contact points available may offer snake robots robustness against unexpected perturbations such as sudden slipping, collisions from other objects, and loss of existing contact. Meanwhile, contact forces at multiple contact points must be coordinated to generate propulsion along desired directions and balanced to maintain stability. To achieve this, terrain contact force sensing and force feedback controllers (Fu and Li, 2021; Kano and Ishiguro, 2020; Liljeback et al., 2014; Ramesh et al., 2021) are needed to sense and adaptively control body bending to maintain contact with the terrain. Snake robots with terrain force sensors and feedback controllers can also be used as robophysical models (Aguilar et al., 2016) to study the mechano-sensory feedback principles. For example, force measurement collected while systematically varying bending strategies can help understand how shape changes are related to contact changes (Fu and Li, 2021; Ramesh et al., 2021). A combination of centralized and decentralized controllers can be tested to study whether and how animal may use similar control mechanisms in the spinal cord to generate complex, robust locomotion patterns (Thandiackal et al., 2021).

**Acknowledgements**

The authors would like to thank Qihan Xuan for discussion, Kaiwen Wang and Daniel Deng for the help with preliminary experiments, and Divya Ramesh for help with animal care. This work was supported by a Burroughs Wellcome Fund Career Award at the Scientific Interface, an Arnold & Mabel Beckman Foundation Beckman Young Investigator award, and the Johns Hopkins University Whiting School of Engineering start-up funds to C.L.

**Competing interests**

The authors declare no conflict of interest.

**Author contributions**

Conceptualization: H.C.A., C.L.; Methodology: Q.F., C.L.; Software: Q.F.; Validation: Q.F., C.L.; Formal analysis: Q.F.; Investigation: Q.F.; Resources: Q.F., C.L.; Data curation: Q.F.; Writing - original draft: Q.F., C.L.; Writing - review and editing: Q.F., H.C.A., C.L.; Visualization: Q.F.; Supervision: C.L.; Project administration: C.L.; Funding acquisition: C.L.

## Supplementary information

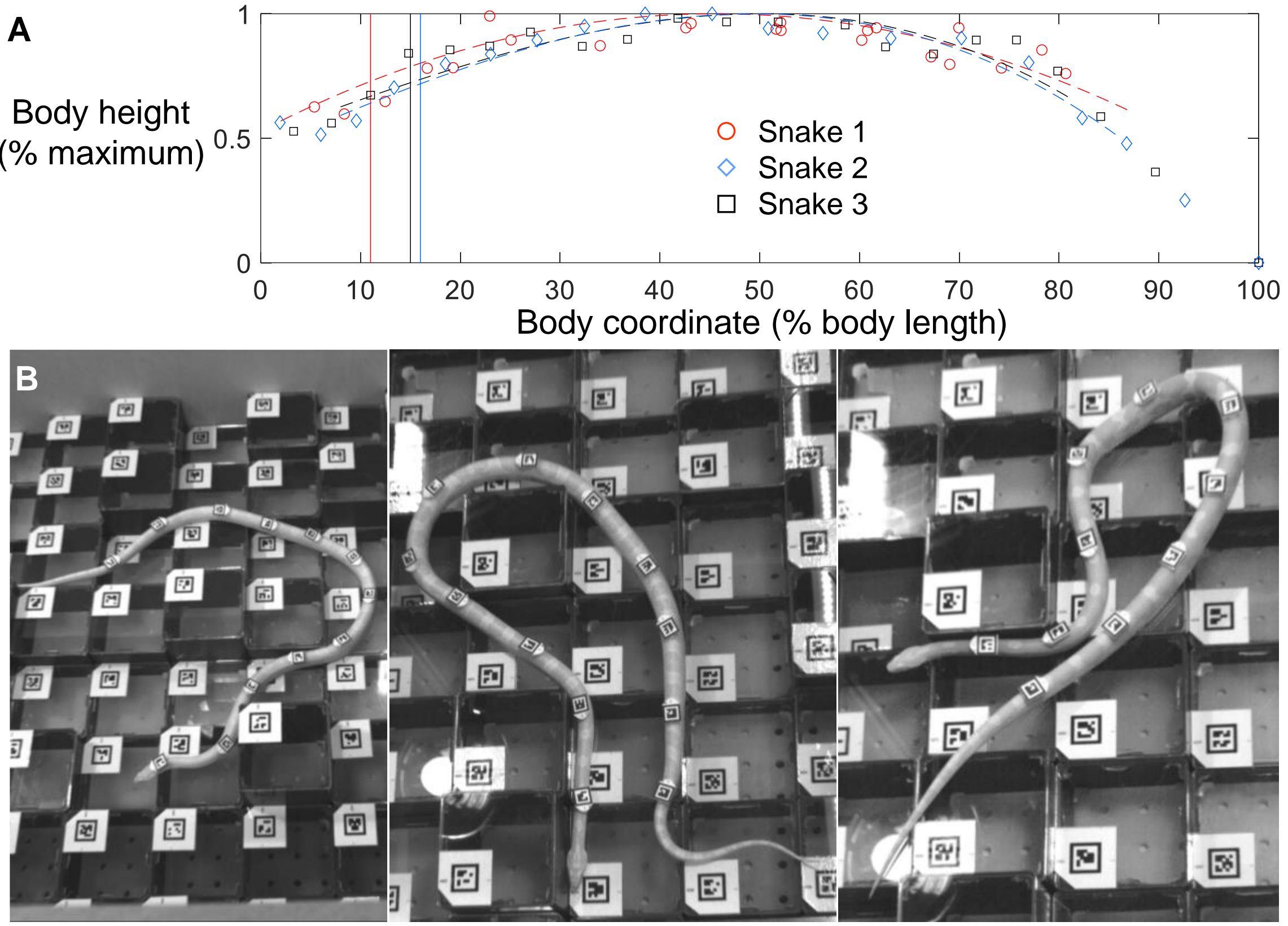

**Fig. S1. Body height and marker distribution along the body. (A)** Measured and fitted height distribution along the body for each individual. Markers show measured height, dashed lines show fitted quadratic polynomials for each individual with corresponding color (trimmed to start from the most anterior marker to the most posterior marker on the body), solid lines show division between the anterior region and the main body region. **(B)** Photos of animals with BEEtag markers attached to the body. Between 10 and 12 BEEtag markers are attached along the dorsal side of the snake equally spaced between neck and vent,

covering an average of 79% of full body length (96% of full body volume) as estimated from tapering data in (A).

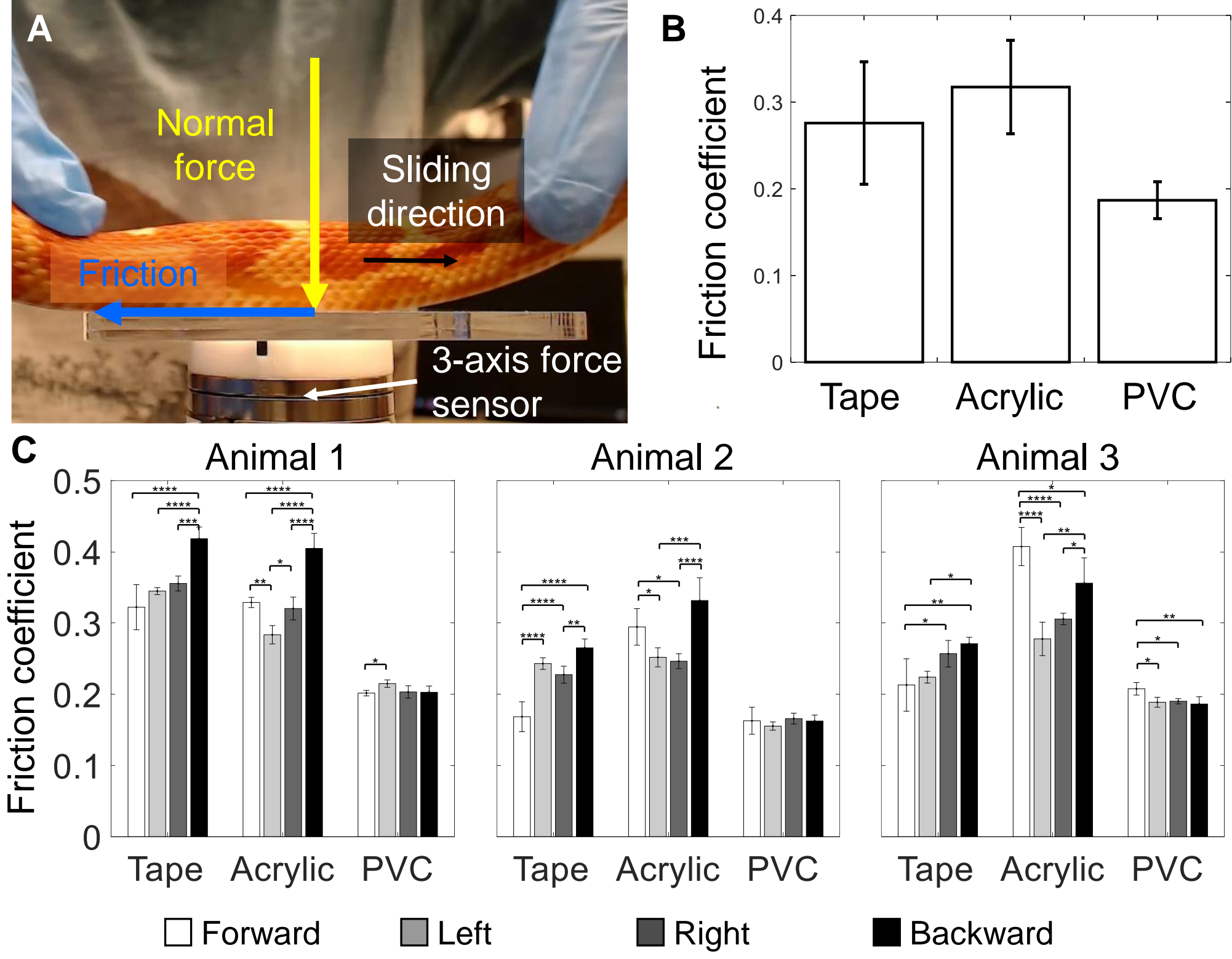

**Fig. S2. Measurement of kinetic friction coefficients. (A)** Setup to measure kinetic friction coefficients between snake body and terrain surfaces. A 3-axis force sensor measures normal force and friction applied to plate while a snake is sliding against plate. Top surface of the plate is covered by acrylic, packaging tape, or PVC in different measurements. **(B-C)** Measured kinetic friction coefficient between snake body and tape, acrylic, and PVC, plotted after pooling all individuals and directions (B) and separately for each individual and each direction (C). Error bars show ± 1 s.d. Brackets and asterisks represent statistically significant differences between two directions (*$P < 0.05$; **$P < 0.005$; ***$P < 0.0005$; ****$P < 0.0001$, ANOVA, Tukey HSD).

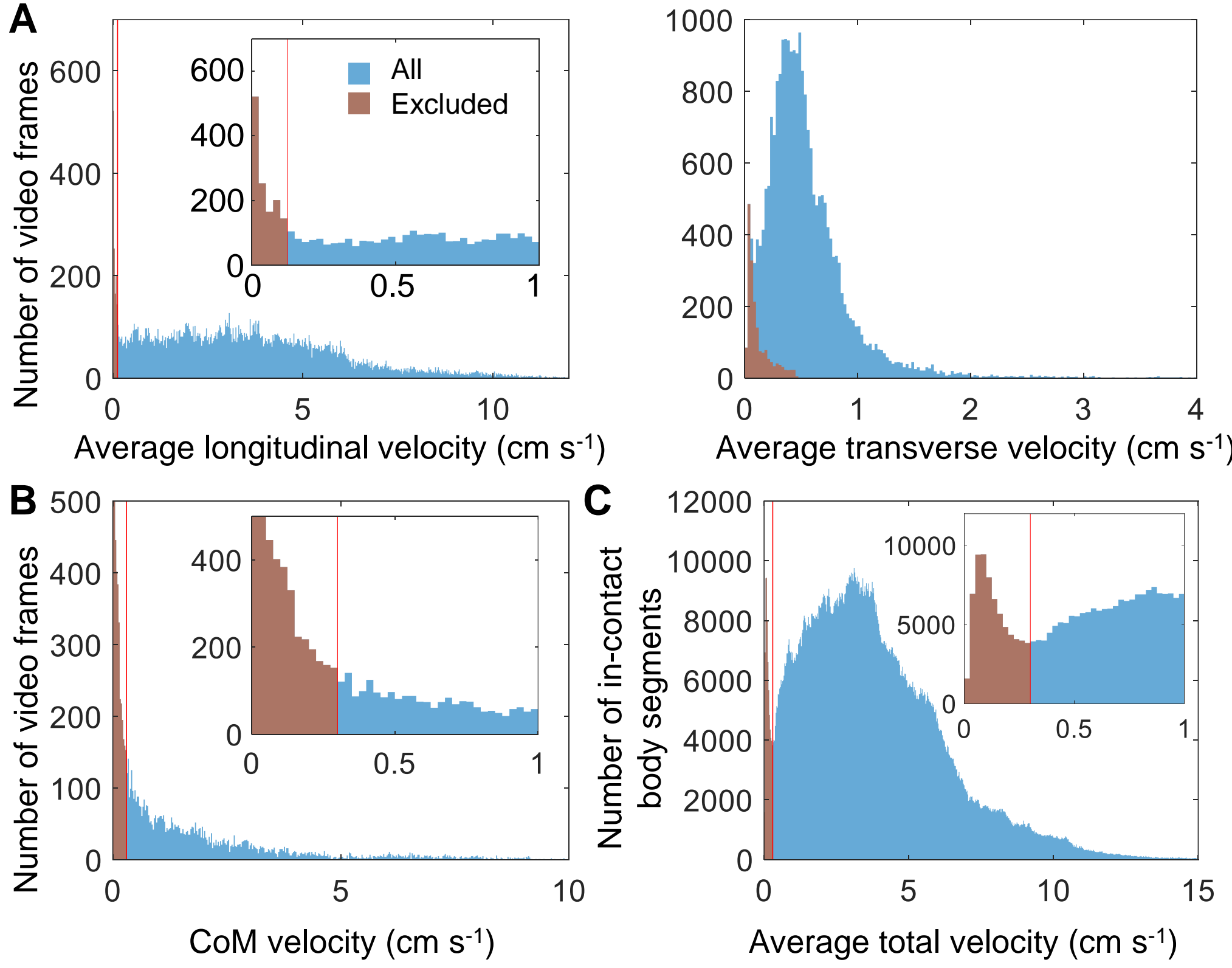

**Fig. S3. Histograms of velocities used when selecting thresholds to exclude data. (A)** Histograms of average longitudinal (left) and average transverse (right) velocities (averaged along the entire reconstructed body) in all video frames pooled from all trials of all individuals (blue) and the excluded phases in which the snake stops (i.e., video frames with a small (< 0.125 cm $s^{-1}$) average longitudinal velocity) (brown). **(B)** Histogram of center of mass (CoM) velocity in all video frames (excluding phases in which the snake stops) pooled from all trials of all individuals (blue) and the excluded video frames in which CoM velocity is small (< 0.3 cm $s^{-1}$) (brown). **(C)** Histogram of average total velocity of the body segments in contact with the terrain pooled from all video frames (excluding phases in which the snake stops or CoM velocity is small) in all trials of all individuals (blue) and the excluded body segments in contact with the terrain which has a small (< 0.3 cm $s^{-1}$) total velocity (brown). Red line shows thresholds of exclusion. Inset shows close-up views at small velocities.

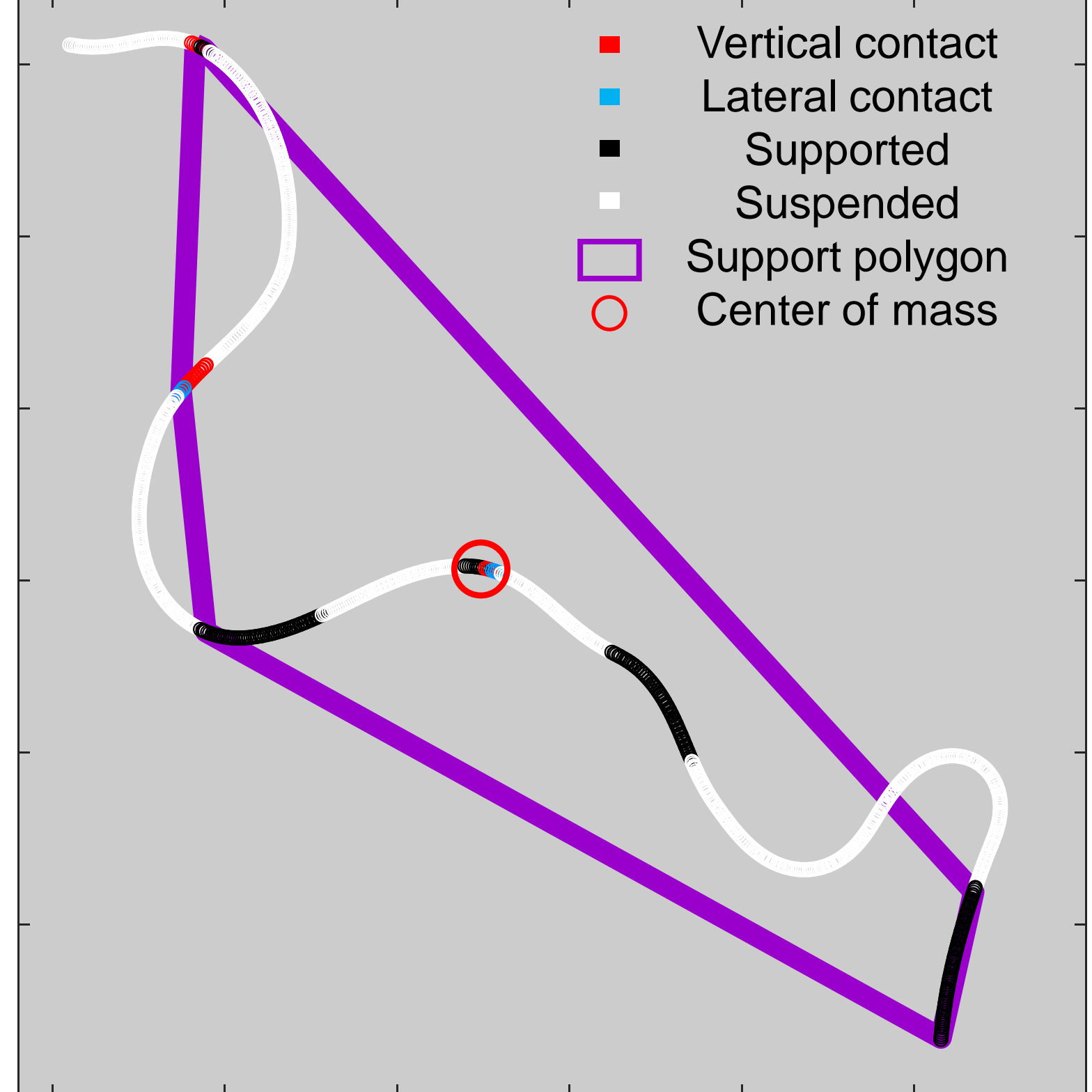

**Fig. S4. Static stability analysis.** Thick curve shows body segments color-coded by contact types. Purple polygon shows support polygon, a convex region formed by body segments in contact with horizontal surfaces. Red circle shows center of mass. When center of mass is inside support polygon, the snake is statically stable.

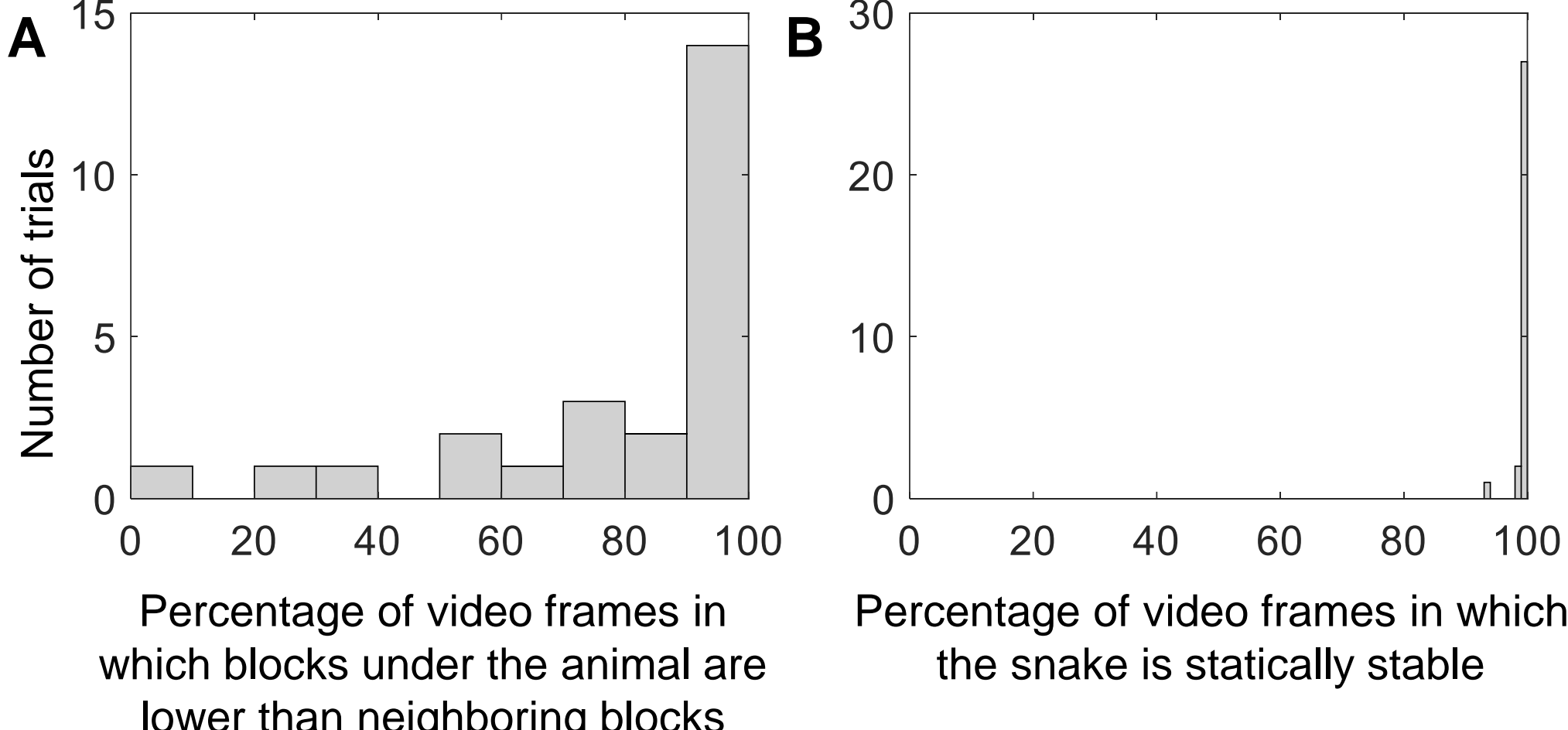

**Fig. S5. Histograms of percentage of video frames in which the snake tends to move on lower blocks and in which the snake is statically stable. (A)** Histogram of the percentage of video frames in which the average height of blocks under the animal body is smaller than that of neighboring blocks. **(B)** Histogram of the percentage of video frames in which the center of mass falls within the support polygon formed by body segments in contact with horizontal surfaces (i.e., when the snake is statically stable).

**Movie 1. A representative trial of a snake utilizing vertical bending to traverse an uneven terrain.**

**Movie 2. A representative trial showing little transverse motion with reconstructed midline overlaid at different time instances.**